\documentclass[aip, pof,
 amsmath,
 amssymb,
 reprint,
showpacs,
showkeys,
]{revtex4-1}
\usepackage{soul, color}
\usepackage{graphicx}
\usepackage{dcolumn}
\usepackage{bm}
\DeclareMathAlphabet{\mathpzc}{OT1}{pzc}{m}{it}

\begin{document}
\title{On the self-similarity of diffracting gaseous detonations and the critical channel width problem}
\author{M. I. Radulescu}
   \email{matei@uottawa.ca}
\affiliation{Department of Mechanical Engineering, University of Ottawa, Ottawa (ON) K1N 6N5 Canada}
\author{R. Mével}
\affiliation{Center for Combustion Energy, School of Vehicle and Mobility, State Key Laboratory for Automotive Safety and Energy, Tsinghua University, Beijing, China}
\author{Q. Xiao}
\affiliation{National Key Laboratory of Transient Physics, Nanjing University of Science and Technology, Nanjing, 210094, China}
\author{S. Gallier}
\affiliation{ArianeGroup, Le Bouchet Research Center, 91710 Vert le Petit, France}
\date{\today}

\begin{abstract}
One strategy for arresting propagating detonation waves in pipes is by imposing a sudden area enlargement, which provides a rapid lateral divergence of the gases in the reaction zone and attenuates the leading shock.  For sufficiently small tube diameter, the detonation decays to a deflagration and the shock decays to negligible strengths.  This is known as the \textit{critical tube diameter} problem.  In the present study, we provide a closed form model to predict the detonation quenching for 2D channels.  Whitham’s geometric shock dynamics, coupled with a shock evolution law based on shocks sustained by a constant source obtained by the shock change equations of Radulescu, is shown to capture the lateral shock dynamics response to the failure wave originating at the expansion corner.  A criterion for successful detonation transmission to open space is that the lateral strain rate provided by the failure wave not exceed the critical strain rate of steady curved detonations.  Using the critical lateral strain rate obtained by He and Clavin, a closed form solution is obtained for the critical channel opening permitting detonation transmission.  The predicted critical channel width is found in very good agreement with our recent experiments and simulations of diffracting H$_2$/O$_2$/Ar detonations.  
\end{abstract}

\maketitle

\section{Introduction}
When a detonation wave emerges from a tube or channel into an open space, the sudden expansion of the gases in the reaction zone of the detonation wave decouples the reaction zone from the shock near the corner\cite{dremin}.  If the opening is sufficiently small, the detonation can completely degenerate into a weak shock followed by a deflagration wave.  This so-called critical tube diameter problem, or critical channel height in 2D, has attracted much interest since the pioneering work of \citet{zeldovich1956experimental}, due to its practical importance in detonation initiation or quenching applications.  For gaseous detonations, a review of the state of the art can be found in the PhD thesis of \citet{schultz2000detonation} for work prior to 2000, while a discussion of the more recent litterature appears in the notable contributions of \citet{ jones2000numerical, meredith2010detonation,sorin2009detonation, khasainov2005detonation, nagura2013detonation, li2016numerical, shi2020reinitiation, gallier2017detonation, pintgen2009detonation, yuan2021computational, li2021influences}.  While this list is not exhaustive, the scope of our present communication is not to provide a critical review of the vast available litterature on the subject.  We note however that in spite of numerous efforts in modeling the diffraction process, a predictive model for the critical tube diameter is still lacking \cite{lee1984dynamic, arienti2003numerical, lee2008detonation}.  The prime objective of the present communication is to formulate such a model.

The lack of a predictive model is due in part to the presence of a cellular structure of the front of all detonations, which modifies the reaction zone structure of detonations and its sensitivity to the global expansion during the diffraction process, as recently shown by \citet{xiao2020role}. For example, a unique signature of the transverse wave structure of detonations is the anomalous scaling of the diffraction process between the 2D channel problem and 3D axisymmetric problem, as first noted by \citet{radulescu2003propagation}  when reviewing previous data, and subsequently confirmed experimnetally by \citet{ meredith2010detonation}. This anomalous scaling of a tube diameter to channel width greater than 2 cannot be reconciled with curvature based criteria for failure and extensions to include non-steady effects\cite{bdzil1986time}, and is beleived to be a unique feature of unstable cellular detonations.  Recovering the correct scaling can be deemed the ultimate\textit{ acid test} for any theoretical model of diffraction for unstable detonations.  Recently, however, \citet{ meredith2010detonation} and \citet{radulescu2002failure} in porous walled tubes, have shown that only weakly unstable detonations waves do obey the ideal curvature-based scaling, substantiating Lee's claim of the failure mechanism being associated with a critical curvature \cite{lee1996}.  This ideal curvature based scaling was verified numerically by \citet{li2016numerical} for weakly unstable detonations.  The present study addresses whether a curvature based prediction of critical transmission is compatible with experiments in weakly unstable detonations and can quantitatively predict the critical diffraction conditions.

Recent work by \citet{xiao2020dynamics} in slowly enlarging channels has shown that the quasi-steady dynamics of such weakly unstable cellular hydrogen detonations at low pressure, which are characterized by much longer reaction zones as compared to induction zones, can be well captured by the predictions of the ZND model with curvature \cite{xiao2020dynamics} and neglect of the cellular structure.  It is thus of interest to verify whether the diffraction process in these same mixtures can be equally well predicted by neglecting the influence of the cellular structure for these conditions using a critical curvature criterion, as suggested by \citet{lee1996}.  The model we formulate, based on a maximum permissible curvature, is thus meant for weakly unstable detonations.  This is the first step towards predicting the behavior of more unstable detonations, as it can serve as a benchmark for more unstable conditions. 

The prediction of the critical conditions for transmission in open space of a detonation wave requires modeling the dynamics of the diffracting detonations and the distribution of transverse flow strain rate (i.e., curvature times flow speed) behind the lead shock.  Previous studies have shown that the reaction zone decouples from the lead shock behind a failure wave propagating to the axis \cite{dremin, arienti2005numerical}.  The shape of the de-coupled shock wave was shown to be approximately self-similar by \citet{bartlma1986diffraction}, and approximately captured by Whitham's geometrical shock dynamics theory for inert shock waves \cite{whitham1974linear}.  Arienti compared the curvature predicted by the model of Whitham with his computations and found it under-predicted the wave curvature \cite{arienti2003numerical}.  Wescott, Stewart and Bdzil (henceforth WSB) have extended Whitham's characteristic rule to under-driven detonation waves by assuming quasi-steady dynamics and a reaction zone with an embedded sonic surface, but did not compare their predictions to gaseous diffracting detonations with sensible shock sensitivity permitting local extinction \cite{wescott2004self}.   Recently, we have conducted detailed numerical simulations and experiments of diffracting detonations in weakly unstable 2H$_2$+O$_2$+2Ar detonations \cite{mevel2017hydrogenP}.  In this study, we wish to compare the predictions of Whitham and WSB curvature distribution behind the diffracting detonation waves with our recent experiments and simulations.    

While both Whitham's and the WSB models will be shown to provide an adequate approximation for the lateral strain rate distribution required for predicting failure, we also introduce a novel self-similar approximation for the shock dynamics weakly supported by the motion of the detonation products.  This model provides an improved prediction for the detonation dynamics.   Combining the prediction of lateral strain rate and the maximum strain rate obtained for steady waves \cite{he1994direct, klein1995curved, yao1995normal}, we obtain a closed form expression for the critical channel height for detonation transmission.  We show that the closed form prediction for detonation failure is in excellent agreement with the simulations and good agreement with experiment. 

The paper is organized as follows.  We first briefly review our previous numerical results and experiments.  We then  compare the detonation dynamics with the predictions of Whitham's Geometrical Shock Dynamics using the Whitham and Westcott's truncations utilizing the characteristic rule, as well as our weakly supported shock model. We conclude by formulating the critical diffraction model based on a maximum curvature and its comparison with experiment and simulations.

\section{NUMERICS AND EXPERIMENTS}

Our recent detonation diffraction experiments and simulations in 2H$_2$+O$_2$+2Ar are discussed in detail elsewhere \cite{mevel2017hydrogenP, gallier2017detonation}.  The numerical simulations were performed using a realistic chemical kinetic scheme for hydrogen combustion using AMR (Adaptive Mesh Refinement) with a minimum grid spacing of 7 $\mu$m. This resolution of 7 $\mu$m corresponds to 40 to 90 grid points in the induction length (depending on pressure), which is well above usual recommendations.  Given that low pressure hydrogen detonations have reaction zones longer than the induction zones by an order of magnitude, and half-reaction zone lengths 2-3 times larger than the induction zone length\cite{xiao2020dynamics}, our effective resolution is on the order of 100 to 200 grid points per half reaction length.  For reference, the recently reported state of the art simulations of \citet{shi2020reinitiation} used only 24 grid points per half reaction length, a limitation imposed by the lack of AMR.  We have tested the convergence of the calculations in straight channels for three different resolutions (7, 14 and 28 $\mu$m) in wide channels permitting up to 10 cells and no effects on cell size were observed. The effect of grid resolution on the diffraction behavior was not performed due to limitation in computational resources.

\begin{figure*}
\begin{center}
\includegraphics[width=\linewidth]{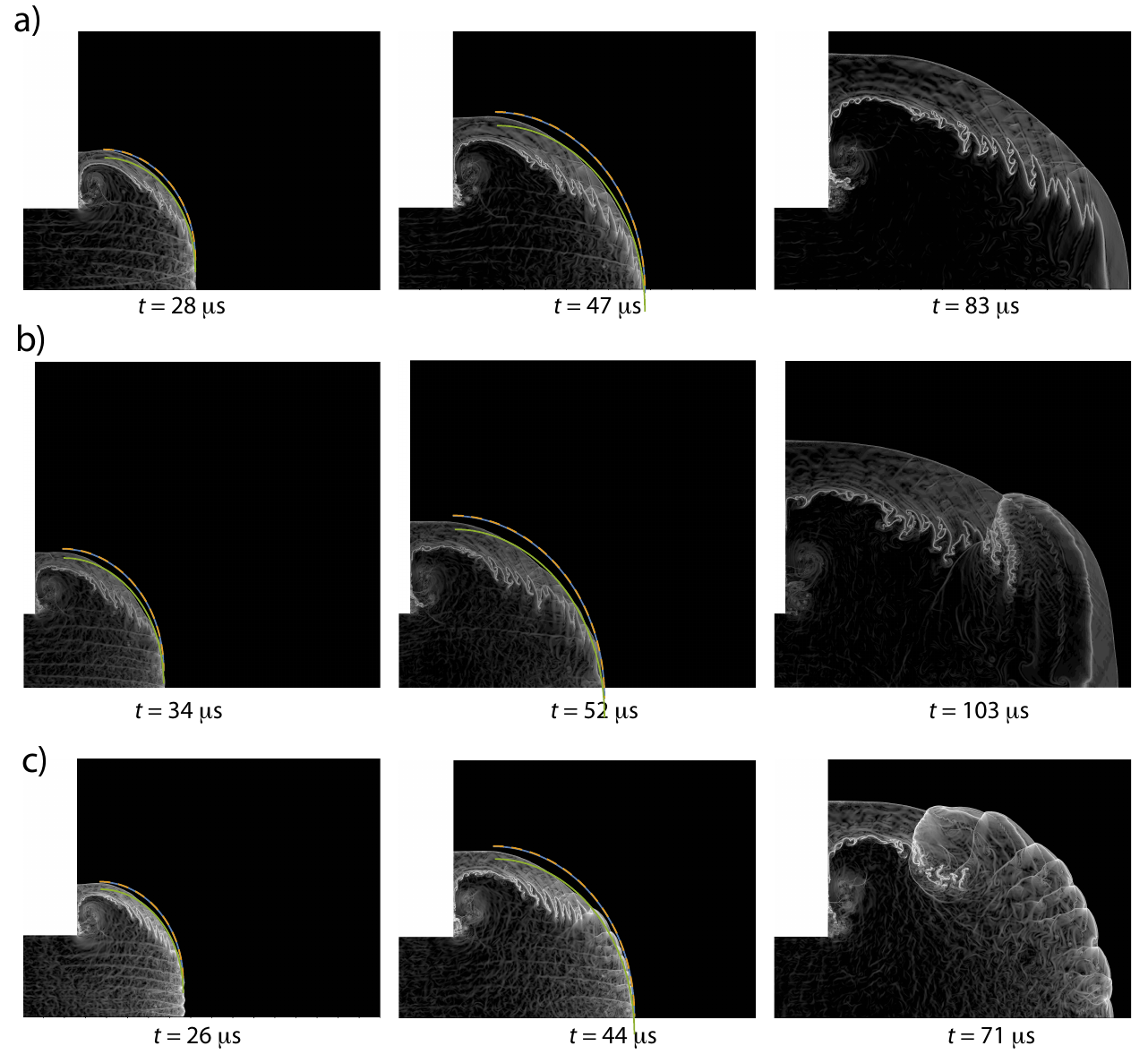} \\
\end{center}
\caption{Numerical Schlieren images adapted from \citet{mevel2017hydrogenP} of diffracting detonations in a 2H$_2$+O$_2$+2Ar mixture at
$T_0=295$K; a) subcritical diffraction, $p_0=6.9$kPa, b) critical diffraction, $p_0=10.3$kPa; c) supercritical diffraction $p_0=13.8$kPa; the height of the computational domain is 188 mm; the time indicated below the images corresponds to the time after the detonation exits the channel; overlaid curves are for a weakly supported shock (green), Whitham's inert shock model (orange) and WSB model (blue); reproduced with permission from the authors.}
\label{fig:numerics}
\end{figure*}

\begin{figure*}
\begin{center}
\includegraphics[width=.8\linewidth]{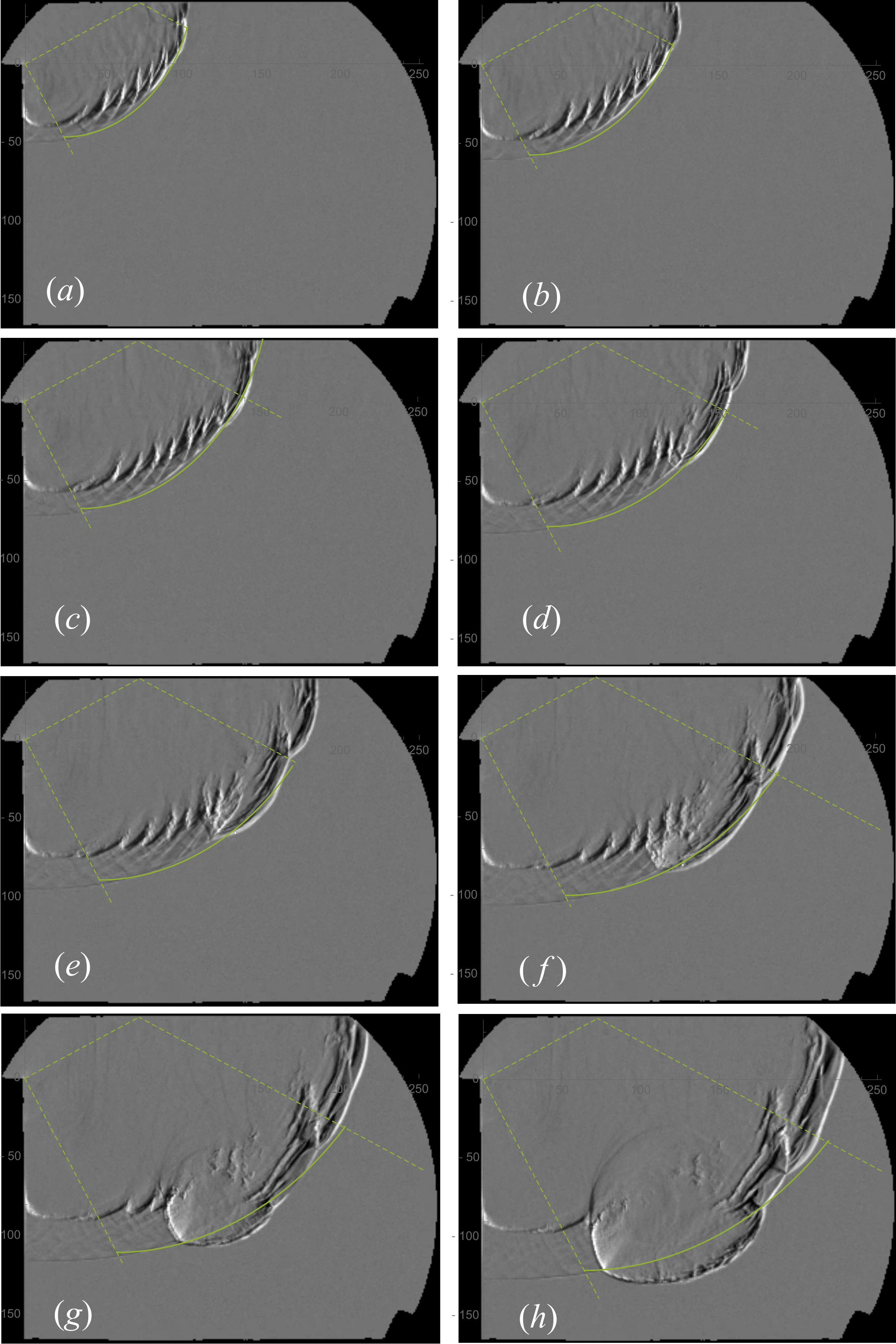} \\
\end{center}
\caption{Composite Schlieren images of detonation diffraction in 2H$_2$+O$_2$+2Ar mixture at
$T_0=295$K and $p_0=17$kPa, adapted from \citet{mevel2017hydrogenP}; the distance between the bottom and top walls is 200 mm; overlaid in green is the shock shape predicted with the weakly supported shock assumption; reproduced with permission from the authors.}
\label{fig:expcritical}
\end{figure*}

Figure \ref{fig:numerics} shows three examples of supercritical, critical and subcritical diffraction obtained in the numerics.  For a fixed geometry, at a sufficiently low pressure and long reaction zone (row a), the detonation wave is quenched by a lateral failure wave originating at the corner, which penetrates to the axis of symmetry.  The shock and reaction zone de-couple and decay.  This is the classical failure dynamics observed in the past experiments and simulations with realistic parameters of gas phase detonations, as also recently documented in detail by \citet{pintgen2009detonation, gallier2017detonation}.  At a sufficiently high pressure and short reaction zone (row c), the failure wave cannot penetrate to the axis and the detonation wave never extinguishes on the axis.  Instead, it continues to propagate with a cellular structure.  At critical conditions intermediate to the two others (row b in the figure), re-inititiation is observed through the amplification of lateral hotspots, previously observed and analyzed by \citet{jones2000numerical, arienti2005numerical, shi2020reinitiation}.  
\begin{figure*}
\begin{center}
\includegraphics[width=0.7\linewidth]{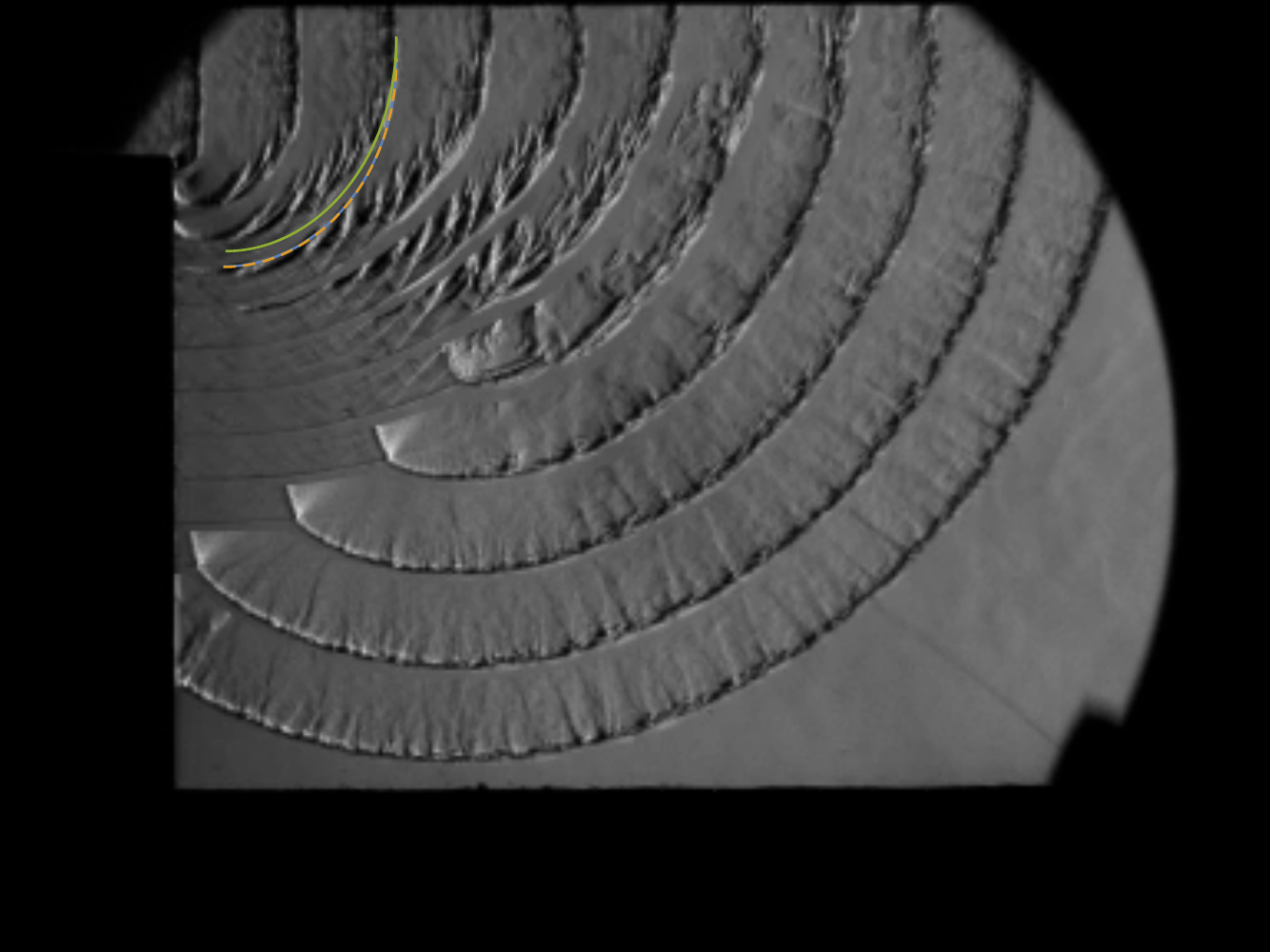} \\
\end{center}
\caption{Composite Schlieren images of detonation diffraction in 2H$_2$+O$_2$+2Ar mixture at
$T_0=295$K and $p_0=23$kPa, adapted from \citet{mevel2017hydrogenP}; the distance between the bottom and top walls is 200 mm; overlaid curves are for a weakly supported shock (green), Whitham's inert shock model (orange) and the WSB model (blue); reproduced with permission from the authors} 
\label{fig:expsuper}
\end{figure*}
The experiments showed very similar dynamics.  Close to the limit, the failure wave does not penetrate to the axis and a curved detonation wave survives quenching (Fig.\ \ref{fig:expsuper}).  At the limit (Fig.\ \ref{fig:expcritical}), re-initiation occurs from a transverse wave amplification into a transverse detonation wave.  The origin of this surviving triple point, identifiable by the protuberance on the failing detonation front, can be traced back to the axis in subfigure a), which also coincides closely to the failure wave arrival at the top wall. In this sense, this critical experiment offers a magnifying glass on the critical dynamics of diffraction.  The detonation wave survives quenching by the amplification of transverse modes over a length scale comparable to the original channel width, signifying that the phenomena controlling sustenance are not strictly local.  The local hotspot formation was analyzed by \citet{jones2000numerical, arienti2005numerical, shi2020reinitiation}.  Of interest, \citet{arienti2003numerical}  further pursued this analysis in the context of Strehlow's ray trapping theory \cite{strehlow1984combustion}.  While this avenue of research appears very worthwhile in explaining the details of the re-initiation mode and the build-up of transverse waves, how the cellular structure impacts the detonation diffraction phenomenon in more unstable mixtures and how potential accompanying turbulent diffusion assists the auto-ignition phenomena \cite{xiao2020role}, this is left for future study.
Table \ref{tab:summary} provides a summary of the limits observed experimentally and numerically in terms of the ratio between the channel bi-thickness $W$ normalized by the ZND induction zone length.  For reducing the experiments, since the detonation prior to diffraction was found to propagate at a speed lower than CJ due to wall losses, the channel width was normalized by the induction zone length calculated for the conditions of the weaker lead shock. The numerical limit of $W_\star/\Delta_i=176$ was found approximately 30\% lower than the experiments, where  $W_\star/\Delta_i$ varied between 200 and 260.  This variability in the experiments is due to the stochasticity, which can be attributed to the cellular structure controlling the details of the limiting phenomena, as illustrated in Fig.\ \ref{fig:expcritical}, for example.\\
\begin{table*}
\caption{Summary of diffraction experiments and model prediction.} \label{tab:summary}
  \centering
  \begin{tabular}{ l c c c c c}
     \hline
    & $n$ in \eqref{eq:form} & $n$ for $\gamma=1.4$ & $W_\star/\Delta_i$ (model) &  $W_\star/\Delta_i$ (num) &  $W_\star/\Delta_i$ (exp)\\
    \hline
    Whitham & $n_W=1+\frac{2}{\gamma}+\sqrt{\frac{2 \gamma}{\gamma-1}}$ & 5.07 & 146 & 176 & 200-260\\
    WSB & $n_{WBS}=3\left( \frac{\gamma+1}{\gamma} \right)$ & 5.14 & 145 &  176 & 200-260\\
   shock support & $n_{R}=2\left( \frac{\gamma+1}{\gamma} \right)$ & 3.43 & 164 &  176 & 200-260\\
   \hline
  \end{tabular}
\end{table*}
\section{WHITHAM'S SELF-SIMILAR GSD SOLUTION FOR THE SHOCK CURVATURE DISTRIBUTION}
\begin{figure}
\begin{center}
\includegraphics[width=\linewidth]{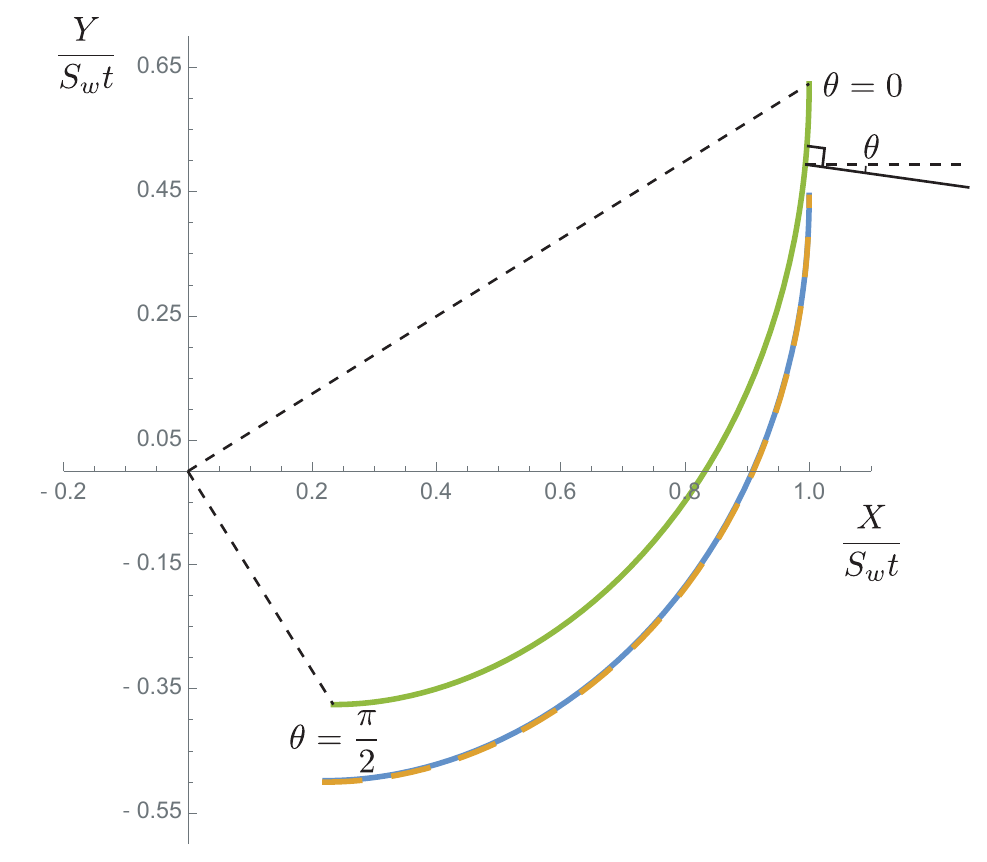} \\
\end{center}
\caption{The self-similar solution predicted by GSD for $\gamma=1.4$:  weakly supported shock (green), Whitham's inert shock model (orange) and WSB model (blue).}
\label{fig:GSD}
\end{figure}

\begin{figure}
\begin{center}
\includegraphics[width=\linewidth]{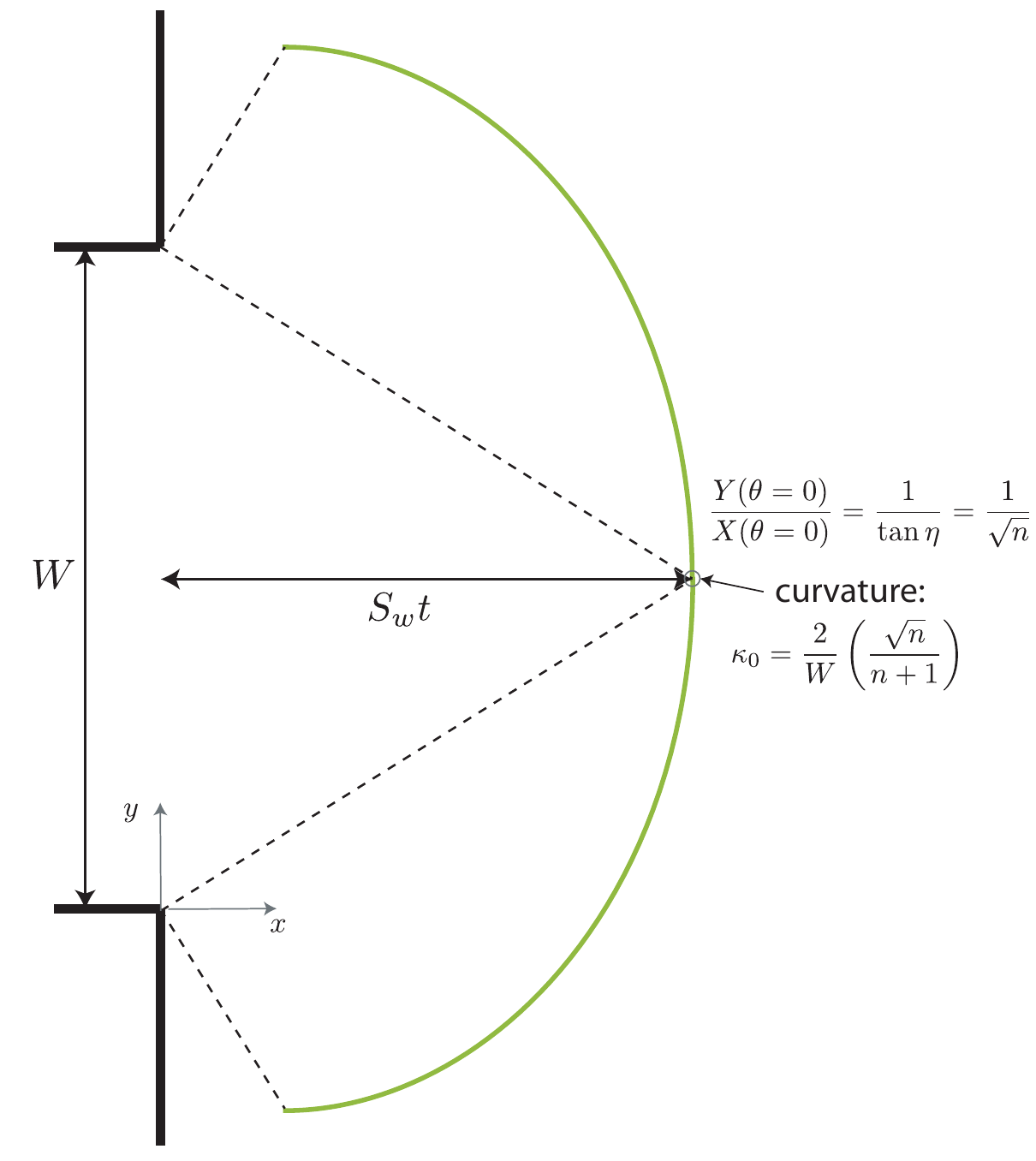} \\
\end{center}
\caption{The GSD construction for approximating the shock shape when the corner signals meet.}
\label{fig:GSD_phys}
\end{figure}

Whitham's geometric shock dynamics (GSD) provides a simple framework to predict the dynamics of shocks affected by changes in the shock inclination angles, such as diffraction, as well as determine the curvature evolution \cite{whitham1974linear}.  For generic local shock evolution equations of the form
\begin{equation}
\frac{S_w^2 \kappa}{\dot{S}_w}\equiv\frac{\mathrm{d} \ln A/\mathrm{d}x}{\mathrm{d} \ln S_w / \mathrm{d}x}=-n \label{eq:form}
\end{equation}
 (i.e., $S_w \propto A^{-1/n}$) where $S_w$ is the normal speed of the shock with respect to a medium at rest,  $\kappa = \frac{\mathrm{d} \ln A}{\mathrm{d}x}$ its local curvature, $A$ a measure of the surface area of a shock element and $n$ an arbitrary exponent, the shape of a diffracting shock over a sharp corner is given by equations (8.95) in Whitham \cite{whitham1974linear}:
\begin{align}
\frac{X}{S_w t} = \sqrt{ \frac{n+1}{n} } \exp\left( \frac{\theta}{\sqrt{n}} \right) \sin \left( \eta-\theta \right)\\
\frac{Y}{S_w t} = \sqrt{ \frac{n+1}{n} } \exp\left( \frac{\theta}{\sqrt{n}} \right) \cos \left( \eta-\theta \right)
\end{align}
where $\eta$ is given by $\tan \eta = \sqrt{n}$ and $\theta$ is the angle of the unit normal to the shock surface with the $x$-axis, see Fig.\ \ref{fig:GSD}.  These expressions derive from a purely geometric theory for how surfaces given by a law like \eqref{eq:form} evolve in space.  The physics are reflected by the exponent $n$, which we treat below. 
Since the shock surface is parameterized by $\theta$ in the form $\Lambda\left(X(\theta,t),Y(\theta,t) \right)$, its local curvature at a given time $t$ is given by  
\begin{equation}
\kappa=\frac{X_{\theta} Y_{\theta\theta} - Y_{\theta} X_{\theta\theta}}{\left(X_{\theta}^2+Y_{\theta}^2\right)^\frac{3}{2}}
\end{equation}
yielding 
\begin{equation}
\kappa=\frac{1}{S_w t} \left( \frac{n}{n+1} \right) \exp\left(- \frac{\theta}{\sqrt{n}} \right) 
\end{equation}
The curvature thus decays with time and with increasing $\theta$.  The maximum curvature occurs for $\theta=0$:
\begin{equation}
\kappa_{0}=\frac{1}{S_w t} \left( \frac{n}{n+1} \right) \label{eq:curvmax}
\end{equation}
Adapting these results for our diffraction problem (see Fig.\ \ref{fig:GSD_phys}), where the half width of the channel is $W/2$, the time required for a transverse signal along the shock to reach $y=W/2$ is $\frac{1}{S_w}\frac{W}{2}\sqrt{n}$.  At this time, the shock curvature given by \eqref{eq:curvmax} becomes
\begin{equation}
\kappa_{0}=\frac{2}{W} \left( \frac{\sqrt{n}}{n+1} \right) \label{eq:curvmaxdif}
\end{equation}
\section{CRITERION FOR DETONATION SUSTENANCE}
For the detonation to survive quenching, the criterion we propose, compatible with our experiments and simulations, is that the maximum curvature of the wave given by \eqref{eq:curvmaxdif} not exceed the critical curvature permitting steady curved detonations.  The analytical work of He and Clavin\cite{he1994direct}, Yao and Stewart\cite{yao1995normal, yao1996dynamics} and Klein et al.\ \cite{klein1995curved} permits to express this critical curvature in closed form.  We use the expression of critical curvature obtained by He and Clavin for quasi-steady square wave detonations with an induction time characterized by an exponential sensitivity on temperature, which was found in very good agreement with realistic chemistry calculations \cite{radulescu2002failure}:
\begin{equation}
\kappa_{\star}^{-1}=\Delta_i \frac{8 {\rm e} }{1-\gamma^{-2}} \left( \frac{E_a}{RT_N} \right) \label{eq:curvhe}
\end{equation}
where $\Delta_i$ is the induction zone length of the CJ detonation, $T_N$ is the temperature behind the shock of the CJ detonation and $\frac{E_a}{RT_N} $ is the non-dimensional activation energy characterizing the sensitivity of the induction time to temperature.  From \eqref{eq:curvmaxdif}, we obtain the critical channel height for successful detonation diffraction: 
\begin{equation}
\frac{W_\star}{\Delta_i}=\frac{2 \sqrt{n}}{n+1}  \frac{8 {\rm e} }{1-\gamma^{-2}} \left( \frac{E_a}{RT_N} \right) \label{eq:Wstar}
\end{equation}
The exponent $n$ depends on the model of shock evolution adopted, as we will see next.
\section{THE EXPONENT $n$ FROM SHOCK EVOLUTION EQUATIONS}
\begin{figure}
\begin{center}
\includegraphics[width=1\linewidth]{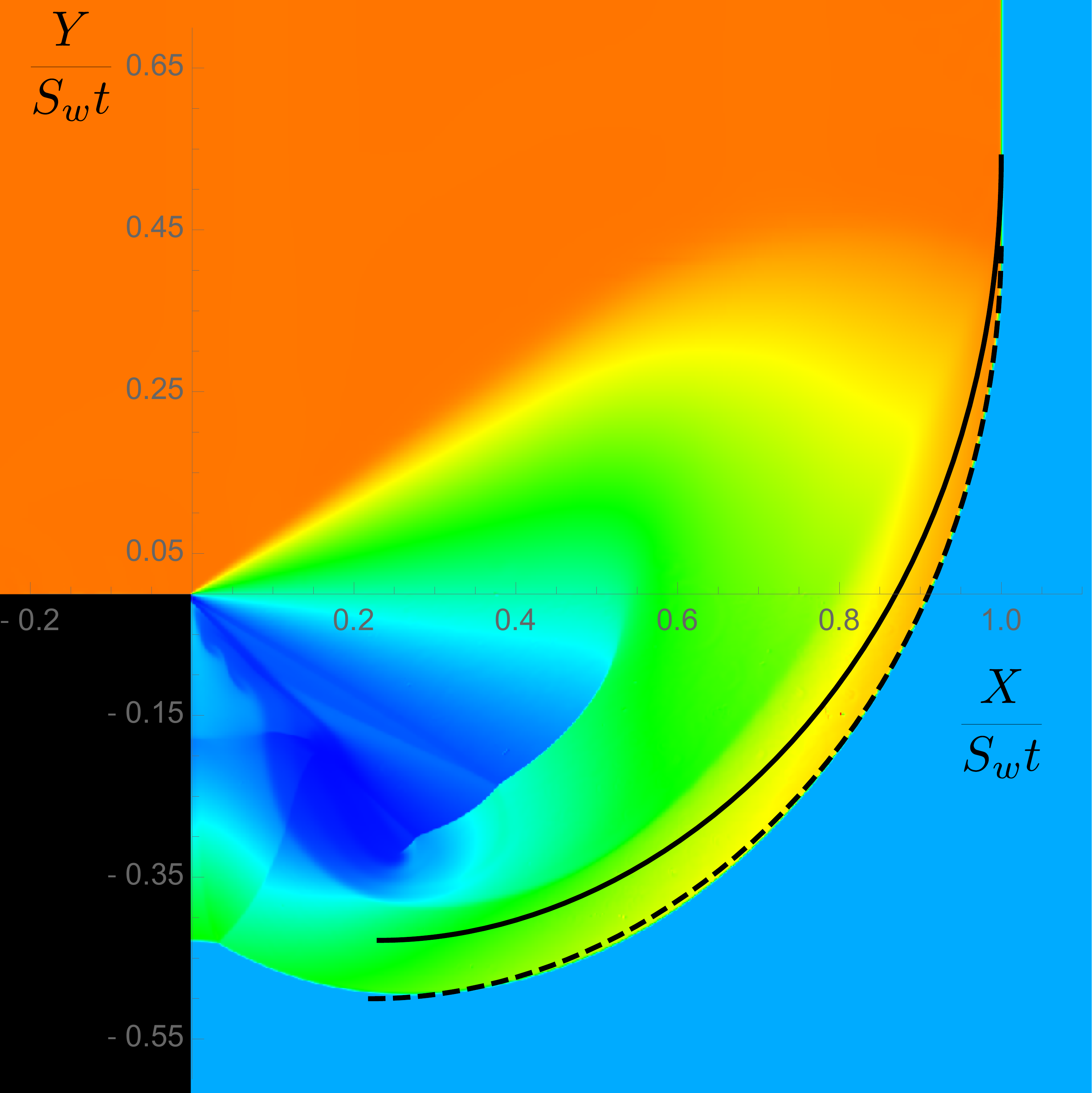} \\
\end{center}
\caption{The density distribution during the diffraction of a $M_w=5$ and $\gamma=1.4$ inert shock at a $\pi/2$ corner; overlaid curves are for a weakly supported shock (solid line) and Whitham's inert shock model obtained with the characteristic rule.}
\label{fig:inert}
\end{figure}

An exact solution for decaying shock waves yielding the exponent $n$ is not currently available.  Physically based approximations have been proposed, and we briefly review those relevant for the present problem of diffraction.
Whitham's model for the shock evolution equation is obtained by projecting the shock state changes along the trajectory of a $C+$ characteristic.  This model applies to shock waves for which the rear boundary conditions play a negligible influence on the shock dynamics.  For strong shocks, Whitham \cite{whitham1974linear} obtains the exponent $n$ given by:
\begin{equation}
n=n_W=1+\frac{2}{\gamma}+\sqrt{\frac{2 \gamma}{\gamma-1}}
\end{equation}
For $\gamma=1.4$, corresponding to the post shock state in the experiments and numerics discussed above, this exponent is 5.07.  The predictions of the decoupled detonation are shown in the Figs.\  \ref{fig:numerics}, \ref{fig:expsuper} and  \ref{fig:expcritical} as the orange curves.  The model is in fair agreement with the experiments and numerics.  Note that the Whitham exponent is sometimes erroneously quoted with $+$ in the square root instead of the $-$.  This unfortunate typographical error in Bartlma and Schroder \cite{bartlma1986diffraction} has persisted in the more modern literature \cite{yuan2020model}.  Using the incorrect exponent leads to better agreement with simulations and experiments, but a purely fortuitous one.  
The characteristic rule has also been applied to underdriven detonations relevant to diffraction problems by Wescott et al.\ \cite{wescott2004self}.  Their model assumes the detonation in quasi-steady state, and requires an embedded sonic surface.  The resulting exponent $n$ of their model for sonic under-driven detonations is:
\begin{equation}
n=n_{WBS}=3\left( \frac{\gamma+1}{\gamma} \right)
\end{equation}
The model reproduces one of the limits obtained by more rigorous perturbation methods by Yao and Stewart\cite{yao1996dynamics}. Again, this exponent has unfortunately also been reported erroneously (as the inverse of this expression above) in the paper of Wescott et al.  For $\gamma=1.4$, corresponding to the post shock state in the experiments and numerics discussed above, the correct exponent is 5.14, i.e., almost identical to the inert shock model of Whitham.  The shock prediction using this model is shown in blue; it provides the same fair agreement as the Whitham model for this value of $\gamma$, although they model fundamentally different phenomena. 
The characteristic rule adopted in the previous two models is aimed to model the dynamics of shock waves not influenced by rear boundary conditions.  This is physically inconsistent with the dynamics observed in experiments.  For the detonation diffraction problem, the arrival of the failure wave quenches the chemical reactions by sudden flow expansion.  The shock is subsequently partially supported by the motion of the products, which act as a piston and influence the shock dynamics, at least in the region immediately adjacent to the arrival of the failure wave and conducive to reaction quenching.  This situation was analyzed for pulsed sources by Chekmarev \cite{chekmarev1975} and Radulescu and Law \cite{radulescu2007transient} in the context of jets issuing from finite sized sources.  These authors have found two asymptotic behaviors, the near field dynamics of the shock were controlled by the source outflow speed, while the far field was controlled by the free dynamics of the mass layer bounded by inner and outer facing shocks.  The early dynamics of the failed detonation fronts can be argued to correspond to the former class.  In the detonation problem, a rear facing shock is not formed, while for the diffraction of a purely inert shock, the rear facing shock is formed to match the post shock state to the supersonic expansion.  Indeed, the flow in the lab frame following a detonation is subsonic, whereas it is supersonic for a strong inert shock.  Based on these physical considerations, a physical model for shock dynamics in supported shocks is to assume that there is no time variation in the piston support of the products, mimicking a constant rear support.  
Radulescu has recently derived an exact expression linking $\frac{\partial u}{\partial t}$ immediately behind the shock with the local dynamics of curved shocks using Fickett and Davis' shock change equations \cite{radulescu2020shock}.  For strong inert shocks, this local balance can be written as
\begin{equation}
\frac{(\gamma+1)^2}{4\gamma} \frac{1}{\dot{S}_w}\frac{\partial u}{\partial t}=2\left( \frac{\gamma+1}{\gamma} \right)+\frac{S_w^2 \kappa}{\dot{S}_w} \label{eq:14}
\end{equation}
To model the weak support provided by the detonation products upon arrival of the failure wave, we assume the evolution of the rear support to vary slower than the shock wave response to area changes, i.e., we neglect the term on the left-hand-side of this equation.  This resulting local quasi-steady approximation for the motion of the piston leads to a simple evolution equation for the shock of the desired form \eqref{eq:form}, with $n$ given by 
\begin{equation}
n=n_{R}=2\left( \frac{\gamma+1}{\gamma} \right)
\end{equation}
For $\gamma=1.4$, corresponding to the post shock state in the experiments and numerics discussed above, this exponent is 3.43.  The shock dynamics predicted by this truncation for the shock dynamics are shown in green in Figs. \ref{fig:numerics}, \ref{fig:expsuper} and  \ref{fig:expcritical}.  The simple model is found in very good  agreement with both the simulations and experiment.  It can thus serve to evaluate the lateral strain rate behind the shock when the failure wave reaches the axis, as discussed above.  We note that the truncation suggested was not derived through a rational derivation using a systematic multi-scale approach.  We note that Whitham's model and the WSB model, and more generally all truncations of this type, are physically justified but do not derive systematically from first principles.  Their good performance in shock and detonation diffraction applications clearly warrants a future effort to derive these approximations from first principles.  For its further numerical validation, future work should also attempt to evaluate each of the terms in \eqref{eq:14} in a ray tube perpendicular to the mean shock locus in order to evaluate the accuracy of the balance suggested.  This would necessitate taking ensemble averages of multiple realizations for achieving a coarse-grained description at a hydrodynamic scale, as attempted in the past in stationary problems by \citet{radulescu2007hydrodynamic} and \citet{reynaud2017}. The non-stationarity of the present problem makes this exercise much more difficult, and it is left for future study.
To further illustrate however how the detonation diffraction problem differs from the problem of diffraction of an inert shock, we have computed numerically the diffraction of a Mach $M=5$ shock for $\gamma=1.4$, the same parameters as for the reactive case.  The self-similar density field is shown in Fig.\ \ref{fig:inert}, along with the predictions of shock dynamics discussed above.  Clearly, for the inert shock dynamics, the shock shape is well captured by the Whitham characteristic rule.  Note however the  barrel shock system that separates the supersonic steady expansion of the flow passing the corner and the gas layered between the inner and outer facing shocks, whose dynamics dictate the shock evolution.  This is fundamentally different from the detonation diffraction process, where these elements are absent and are replaced by a wall vortex.  
The discussion of the shock dynamics predicted by different physics has shown that the shape of the diffracting shock is weakly dependent on which model is used, since the conditions immediately behind the shock affect its motion.  More fundamentally, while the diffraction process is not expected to be strictly self-similar, as the local front is more likely well approximated by the WSB model close to the axis where the reactions are still coupled, our model when reactions decouple and the shock is weakly supported by the motion of the detonation products, and the Whitham inert model further from the axis for free-decay, the close coincidence of these exponents suggests quasi-self similarity.  For engineering purposes, taking any of them should be satisfactory. Nevertheless, a formal multi-scale analysis is warranted in the framework of the theory of Detonation Shock Dynamics of \citet{bdzil1986time} formulated for a kinetic law compatible with the long reaction zone character of hydrogen detonations and activated induction zone, such as the model investigated by \citet{short_bdzil_2003}.  This is well outside the scope of the present discussion, which aims at formulating an approximate transmission criterion. 
\section{Transmission criterion}
With the exponent $n$ in closed form from the different approximations discussed above, the critical channel height for detonation transmission to open space can be obtained in closed form by substituting the expressions for $n$ given above in \eqref{eq:Wstar}.  For our model with quasi-constant rear support, negligible non-steady effects and negligible transverse wave structure, in the limit of a square wave reaction zone structure, we obtain:
\begin{equation}
\frac{W_\star}{\Delta_i}=  \frac{16 \sqrt{2}  {\rm e} \sqrt{\gamma^5 (\gamma+1)} } {(2+3\gamma)(\gamma^{2}-1)} \left( \frac{E_a}{RT_N} \right) \label{eq:Wstarfinal}
\end{equation} 
Similar expressions can be obtained using the other expressions for the constant $n$.  For the 2H$_2$+O$_2$+2Ar mixture tested, the postshock $\gamma$ is 1.4 and the reduced activation energy is 4.4, obtained from the sensitivity of the ignition delay to temperature changes at the Von Neuman state using Cantera for the calculations.  The resulting predictions of critical channel width using the different values of $n$ are shown in Table 1, along with the experimental and numerical values obtained in our previous experiments and simulations discussed above.  The weakly supported shock model predicts the critical value obtained from the simulations with an error of less than 7\%.  The predictions based on Whitham and Westcott et al. models for the shock dynamics underestimate the critical channel width by 17\%, which is also quite remarkable.  All the models underpredict the experiment, for which the limit is approximately 30\% larger than for the numerics.
\section{CONCLUSIONS}
The proposed model for critical detonation diffraction relies on the prediction of the wave curvature effected by the failure wave originating from the diffraction corner.  While the models of Whitham and Westcott, Bdzil and Stewart are found to predict these dynamics fairly well, we propose an improvement on the shock dynamic prediction using a weakly supported shock model, which is found in very good agreement with experiment and simulations.  Using these simple estimates for the maximum wave curvature attained when the failure waves meet the axis, a simple criterion for successful transmission is that this curvature not exceed the maximum curvature that can be sustained by a curved detonation in quasi-steady state.  The closed form limits obtained for the critical diffraction channel height are found in very good agreement with numerics and under-predict the experiments by approximately 40\%.  While the present study suggests that the diffraction of weakly unstable detonations are well predicted by the maximum curvature of a steady ZND detonation, future work should address how this prediction performs in much more unstable detonations, where non-steady effects become more prominent\cite{austin} and the turbulent structure of the reaction zone provides a mechanism to enhance the ignition process\cite{xiao2020role}.  Indeed, detonation attenuation with controlled side relief in porous walled tubes \cite{radulescu2002failure, radulescu2003propagation} and weakly divergent tubes \cite{xiao2020role} showed that the curved ZND model predictions fail with increasing instability.    
\section*{Acknowledgements} 
MIR acknowledges financial support provided by the Natural Sciences and Engineering Research Council of Canada (NSERC) through the Discovery Grant ”Predictability of detonation wave dynamics in gases: experiment and model development”.   
\section*{AVAILABILITY OF DATA}
The data that support the findings of this study are available from the corresponding author upon reasonable request.
\bibliography{references}

\end{document}